\def\OX{\Omega_\textrm{X}}
\def\SX{S_\textrm{X}}
\def\IX{I_\textrm{X}}

\def\aX{a_\textrm{X}}
\def\eX{e_\textrm{X}}
\def\masy{~\textrm{mas~yr}^{-1}}
\def\muasy{~\upmu\textrm{as~yr}^{-1}}
\def\deg{~\textrm{deg}}

\def\kx{{\hat{S}}_x}
\def\ky{{\hat{S}}_y}
\def\kz{{\hat{S}}_z}

\def\nk{n_{\rm b}}

\def\Pb{P_{\rm b}}

\def\rfr#1{Equation~(\ref{#1})}
\def\rfrs#1#2{Equations~(\ref{#1})~and~(\ref{#2})}
\def\Rfr#1{Eq. (\ref{#1})}

\def\derp#1#2{\rp{\partial{#1}}{\partial{#2}}}

\def\eqi{\begin{equation}}
\def\eqf{\end{equation}}
\def\eqia{\begin{eqnarray}}
\def\eqfa{\end{eqnarray}}

\def\rp#1#2{{#1\over#2}}
\def\lb#1{\label{#1}}

\def\bds#1{\boldsymbol{#1}}

\def\ton#1{\left(#1\right)}
\def\qua#1{\left[#1\right]}
\def\grf#1{\left\{#1\right\}}

\documentclass[onecolumn]{aastex}

\usepackage[title]{appendix}
\usepackage{hyperref}
\usepackage{booktabs}
\usepackage[table,xcdraw]{xcolor}
\usepackage{multirow}
\usepackage{rotating,tabularx}
\usepackage{float}
\usepackage{enumerate}
\usepackage[polutonikogreek,english]{babel}
\usepackage{amsmath,starfont,textgreek,w-greek,wasysym}
\usepackage{amsthm}
\usepackage{amscd,lineno}
\usepackage{amssymb,dsfont}
\usepackage{graphicx,epsfig}
\usepackage{txfonts}
\usepackage{xr-hyper}

\RequirePackage{color}

\newcommand{\emaila}{lorenzo.iorio@libero.it}

\allowdisplaybreaks[1]

\begin{document}

\title{A post-Newtonian gravitomagnetic effect on the orbital motion of a test particle around its primary induced by the spin of a distant third body}

\shortauthors{L. Iorio}

\author{Lorenzo Iorio\altaffilmark{1} }
\affil{Ministero dell'Istruzione, dell'Universit\`{a} e della Ricerca
(M.I.U.R.)-Istruzione
\\ Permanent address for correspondence: Viale Unit\`{a} di Italia 68, 70125, Bari (BA),
Italy}

\email{\emaila}

\begin{abstract}
We study a general relativistic gravitomagnetic 3-body effect induced by the spin angular momentum ${\bds S}_\textrm{X}$ of a rotating mass $M_\textrm{X}$ orbited at distance $r_\textrm{X}$ by a local gravitationally bound restricted two-body system $\mathcal{S}$ of size $r\ll r_\textrm{X}$ consisting of a test particle revolving around a massive body $M$. At the lowest post-Newtonian order, we analytically work out the doubly averaged rates of change of the Keplerian orbital elements of the test particle by finding non-vanishing long-term effects for the inclination $I$, the node $\Omega$ and the pericenter $\omega$. Such theoretical results are confirmed by a numerical integration of the equations of motion for a fictitious 3-body system. We numerically calculate the magnitudes of the post-Newtonian gravitomagnetic 3-body precessions for some astronomical scenarios in our solar system. For putative man-made orbiters of the natural moons Enceladus and Europa in the external fields of Saturn and Jupiter, the relativistic precessions due to the angular momenta of the gaseous giant planets can be as large as $\simeq 10-50~\textrm{milliarcseconds~per~year}~\ton{\textrm{mas~yr}^{-1}}$. A preliminary numerical simulation shows that, for certain orbital configurations of a hypothetical Europa orbiter, its range-rate signal $\Delta\dot\rho$ can become larger than the current Doppler accuracy of the existing spacecraft Juno at Jupiter, i.e. $\upsigma_{\dot\rho}=0.015~\textrm{mm~s}^{-1}$, after 1 d. The effects induced by the Sun's angular momentum on artificial probes of Mercury and the Earth are at the level of $\simeq 1-0.1~\textrm{microarcseconds~per~year}~\ton{\upmu\textrm{as~yr}^{-1}}$.
\end{abstract}

keywords{
General relativity and gravitation; Experimental studies of gravity; Experimental tests of gravitational theories; Satellite orbits
}

%

\section{Introduction}\lb{intro}
Let us consider a local gravitationally bound restricted two-body system $\mathcal{S}$ composed by a test particle completing a full orbital revolution around  a planet of mass $M$ at distance $r$ in a time interval $\Pb$, and a distant 3rd body X with mass $M_\textrm{X}\gg M$ and proper spin ${\bds S}_\textrm{X}$ around which $\mathcal{S}$ revolves at distance $r_\textrm{X}\gg r$ with orbital period $\Pb^\textrm{X}$. In general, $M$ may be endowed with its own Newtonian and post-Newtonian mass and spin multipole moments \citep{2016JGeod..90.1345S,2018RSOS....580640F} affecting the satellite's motion with known \citep{2005som..book.....C,2014grav.book.....P} and less known \citep{2014PhRvD..89f4067A,2017FrASS...4...11S,2018CeMDA.130...40S} Newtonian and post-Newtonian orbital effects like the classical oblateness-driven orbital precessions, the gravitoelectric Einstein pericentre shift, the gravitomagnetic Lense-Thirring effect, etc. Let us consider a kinematically rotating and dynamically non-rotating coordinate system $\mathcal{K}$ \citep{1989NCimB.103...63B,1994PhRvD..49..618D,2011rcms.book.....K} attached to $M$ in geodesic motion through the external spacetime deformed by the mass-energy currents of X, assumed stationary in a kinematically and dynamically non-rotating coordinate system $\mathcal{K}_\textrm{X}$ whose axes point towards the distant quasars \citep{1989NCimB.103...63B,1994PhRvD..49..618D,2011rcms.book.....K}. The planetocentric motion of the test particle referred to $\mathcal{K}$ is further affected by two peculiar post-Newtonian 3-body effects: the time-honored De Sitter precession due to solely the mass $M_\textrm{X}$ \citep{1916MNRAS..77..155D,1918KNAB...27.214S,1921KNAB...23..729F}, and a gravitomagnetic shift due to ${\bds S}_\textrm{X}$ which, to our knowledge, has never been explicitly and clearly calculated in the literature, if ever it had been.
Our purpose is to analytically work out the latter effect at the lowest post-Newtonian order without any a-priori simplifying assumptions concerning both the orbital configurations of the planetocentric satellite's motion and the trajectory of the planet-satellite system $\mathcal{S}$ in the external field of X,  and for an arbitrary orientation of ${\bds S}_\textrm{X}$ in space. For previous, approximate calculation restricted to the orbital angular momentum of the Moon orbiting the Earth in the field of the rotating Sun, see
\citet[Sec.~3.3.3]{1992gill}.

The plan of the paper is as follows. In Section~\ref{calc}, we analytically work out the long-term rates of change of the Keplerian orbital elements of the test particle. Section~\ref{encel} is devoted to the application of the obtained results to some astronomical scenarios in our solar system. We summarize our results and offer our conclusions in Section~\ref{conclu}. For the benefit of the reader, Appendix~\ref{appena} contains a list of the definitions of the symbols used in the paper, while their numerical values and tables are collected in Appendix~\ref{appenb}.
\section{The doubly averaged satellite's orbital precessions}\lb{calc}
In the weak-field and slow-motion approximation, the gravitomagnetic 3-body potential induced by the angular momentum ${\bds S}_\textrm{X}$ of the external spinning object X on the planetary satellite is
\eqi
U_\textrm{GM} = \rp{G}{c^2 r_\textrm{X}^3}{\bds{S}}_\textrm{X}\bds\cdot\qua{-\bds L + 3\ton{{\bds L}\bds\cdot{\bds{\hat{r}}}_\textrm{X}  }{\bds{\hat{r}}}_\textrm{X} }\lb{USS}.
\eqf
In \rfr{USS}, $G,~c$ are the Newtonian constant of gravitation and the speed of light in vacuum, respectively, while $\bds L$ is the angular momentum of the test particle's orbital motion around $M$.
\Rfr{USS} comes from Eq.~(2.19) of \citet[p.~155]{1979GReGr..11..149B} for the interaction potential energy $V_{S1,S2}$ of two spins ${\bds S}^{(1)},~{\bds S}^{(2)}$ of masses $m_1,~m_2$ separated by a distance $r$ and moving with relative speed $v$
in the limit $m_2\equiv M_\textrm{X}\gg m_1\equiv M$, and by assuming that the spin ${\bds S}^{(1)}$ is the orbital angular momentum of the planetocentric satellite's motion while ${\bds S}^{(2)}$ is the spin angular momentum ${\bds S}_\textrm{X}$ of the distant 3rd body X. Thus, $r$ in Eq.~(2.19) of \citet[p.~155]{1979GReGr..11..149B} has to be identified with $r_\textrm{X}$, and $\bds r\bds\times \bds P$ is nothing but the orbital angular momentum ${\bds r}_\textrm{X}{\bds \times} M{\bds v}_\textrm{X}$ of the motion of $\mathcal{S}$ around $M_\textrm{X}$.
It is interesting to note that, with the same identifications, $V_{S1}$ and $V_{S2}$ of  Eqs.~(2.17)-(2.18) in \citet[p.~155]{1979GReGr..11..149B} yield the gravitoelectric De Sitter orbital precession for the planetocentric motion of the satellite and the gravitomagnetic Lense-Thirring effect for the X-centric orbit of $M$, respectively.

The perturbing potential $U_\textrm{pert}$ to be inserted into the Lagrange planetary equations for the rates of change of the osculating Keplerian orbital elements of the test particle \citep{Bertotti03,2011rcms.book.....K}, obtained by doubly averaging \rfr{USS} with respect to $P_\textrm{b},~P_\textrm{b}^\textrm{X}$ for arbitrary orbital configurations of both the external body X and the test particle and for a generic orientation of ${\bds S}_\textrm{X}$ in space, is
\eqi
U_\textrm{pert} \lb{pluto} = \overline{\overline{U}}_\textrm{GM} = -\rp{G\SX\nk a^2\sqrt{1-e^2}}{2c^2\aX^3\ton{1 - \eX^2}^{3/2}}\mathcal{U},
\eqf
with
\begin{align}
\mathcal{U}\nonumber & =\cos I \grf{2 \kz -3 \sin\IX \qua{\kz \sin\IX + \cos\IX \ton{\ky \cos\OX - \kx \sin\OX}}} + \\ \nonumber \\
\nonumber &+  \rp{\sin I}{2} \grf{2 \ky \cos\Omega - 2 \kx \sin\Omega + 3 \cos\ton{\Omega - \OX} \qua{\kz \sin 2\IX +\right.\right.\\ \nonumber \\
&+\left.\left.2 \sin^2\IX \ton{-\ky \cos\OX + \kx \sin\OX}}}.\lb{paper}
\end{align}
In \rfrs{pluto}{paper}, $a,~e,~I,~a_\textrm{X},~e_\textrm{X},~I_\textrm{X}$ are the semimajor axes, the eccentricities and the inclinations of the orbits of the test particle and of $\mathcal{S}$, respectively, while $\nk$ is the Keplerian orbital motion of the satellite's planetary motion about $M$.
Equations~(\ref{pluto})~and~(\ref{paper}) were obtained in two steps. First, $U_\textrm{GM}$ of \rfr{USS} was evaluated onto the unperturbed ellipse of the planetocentric satellite motion through the standard Keplerian formulas of the restricted two-body problem (see, e.g., Equations~(3.40a)~to~(3-41c) of \citet{2014grav.book.....P}). Then, it was averaged over one orbital period $\Pb$
to the first order in the disturbing potential by using just the Keplerian part of Equation (3.66) of \citet{2014grav.book.....P} for $df/dt$, where $f$ is the true anomaly. Then, the resulting averaged potential ${\overline{U}}_\textrm{GM}$ was, in turn, calculated onto the unperturbed X-centric Keplerian trajectory of $\mathcal{S}$ and averaged over $P^\textrm{X}_\textrm{b}$ to the first order in the perturbation under consideration, thus finally obtaining the double average of \rfrs{pluto}{paper}.

Inserting \rfrs{pluto}{paper} into the right-hand-sides of the Lagrange planetary equations  allows to calculate the doubly averaged rates of change of the Keplerian orbital elements. They turn out to be
\begin{align}
\dot a & = \lb{adot}0, \\ \nonumber \\
\dot e & = \lb{edot}0, \\ \nonumber \\
\dot I \lb{Idot}& = -\rp{G\SX}{2 \aX^3 c^2\ton{1 - \eX^2}^{3/2}}\mathcal{I}, \\ \nonumber \\
\dot\Omega \lb{Odot}& = -\rp{G\SX}{2 \aX^3 c^2 \ton{1 - \eX^2}^{3/2}}\mathcal{O},\\ \nonumber \\
\dot\omega \lb{odot}& = -\rp{G\SX\csc I}{8 \aX^3 c^2 \ton{1 - \eX^2}^{3/2}}\mathcal{P}.
\end{align}
with
\begin{align}
\mathcal{I} \nonumber & = \sin\Omega \grf{-\ky + 3  \sin\IX \cos\OX \qua{-\kz \cos\IX + \sin\IX \ton{\ky \cos\OX - \kx \sin\OX}}} +\\ \nonumber \\
&+ \cos\Omega \grf{-\kx + 3 \sin\IX \sin\OX \qua{\kz \cos\IX + \sin\IX \ton{-\ky \cos\OX + \kx \sin\OX}}}\lb{Icoef}, \\ \nonumber\\
\mathcal{O} \nonumber & =  2 \kz + \kx \cot I \sin\Omega - 3 \cos\IX \sin\IX \ton{\ky \cos\OX - \kx \sin\OX +
\kz \cot I \sin\Omega \sin\OX} - \\ \nonumber \\
\nonumber &- 3 \sin^2\IX \qua{\kz + \cot I \sin\Omega \sin\OX \ton{-\ky \cos\OX + \kx \sin\OX}} + \\ \nonumber \\
& +\cos\Omega \cot I \grf{-\ky + 3 \sin\IX \cos\OX \qua{-\kz \cos\IX + \sin\IX \ton{\ky \cos\OX - \kx \sin\OX}}},\lb{Ocoef} \\ \nonumber \\
\mathcal{P} \nonumber & = \ky \qua{\cos\Omega - 3\cos\ton{\Omega - 2 \OX}} - \kx \qua{\sin\Omega + 3 \sin\ton{\Omega - 2 \OX}} +\\ \nonumber \\
&+ 6 \cos\ton{\Omega - \OX} \qua{\kz \sin 2\IX + \cos 2\IX \ton{\ky \cos\OX -\kx \sin\OX}}.\lb{ocoef}
\end{align}
We remark that \rfrs{adot}{ocoef} are exact in both $e$ and $e_\textrm{X}$ in the sense that the low-eccentricity approximation was not adopted in the calculation.

A more computationally cumbersome approach to obtain the same long-term rates of change of \rfrs{adot}{ocoef} consists, first of all, in deriving a perturbing acceleration from \rfr{USS}.
By writing the Lagrangian per unit mass of a gravitationally bound restricted two-body system affected by a generic perturbing potential as
\eqi
\mathcal{L} = \mathcal{L}_0 + \mathcal{L}_\textrm{pert} = \rp{v^2}{2} + \rp{\mu}{r} + \mathcal{L}_\textrm{pert},
\eqf
the conjugate momentum per unit mass is, by definition,
\eqi
\bds p \doteq \derp{\mathcal{L}}{\bds v} = \bds v + \derp{\mathcal{L}_\textrm{pert}}{\bds v}.
\eqf
Thus,
\eqi
\dot{\bds p} = \dot{\bds v} + \rp{d}{dt}\ton{\derp{\mathcal{L}_\textrm{pert}}{\bds v}}.\lb{pdot1}
\eqf
The Hamiltonian per unit mass is
\eqi
\mathcal{H} = \mathcal{H}_0 + \mathcal{H}_\textrm{pert} = \rp{v^2}{2} -\rp{\mu}{r} + \mathcal{H}_\textrm{pert}.
\eqf
From the Hamilton equations of motion, it is
\eqi
\dot{\bds p} = -\derp{\mathcal{H}}{\bds r} = -\rp{\mu}{r^3}\bds r - \derp{\mathcal{H}_\textrm{pert}}{\bds r}.\lb{pdot2}
\eqf
Since  $\mathcal{L}_\textrm{pert} = -\mathcal{H}_\textrm{pert}$, by comparing \rfr{pdot1} and \rfr{pdot2}, it turns out that the perturbing acceleration is just
\eqi
{\bds A}^\textrm{pert} = \rp{d}{dt}\ton{\derp{\mathcal{H}_\textrm{pert}}{\bds v}} - \derp{\mathcal{H}_\textrm{pert}}{\bds r}.
\eqf
In our specific case, since $\mathcal{H}_\textrm{pert} = U_\textrm{GM}$, we have
\eqi
{\bds A}^\textrm{GM} = \rp{d}{dt}\ton{\derp{U_\textrm{GM}}{\bds v}} - \derp{U_\textrm{GM}}{\bds r} =  \rp{2G}{c^2 r^3_\textrm{X}}\bds v\bds\times\qua{{\bds{S}}_\textrm{X} -3\ton{{\bds S}_\textrm{X}\bds\cdot {\bds{\hat{r}}}_\textrm{X}  }{\bds{\hat{r}}}_\textrm{X} }.\lb{ASS}
\eqf
Then, \rfr{ASS} must be decomposed into its radial ($R$), transverse ($T$) and out-of-plane ($N$) components, which are
\begin{align}
A_R^\textrm{GM} \nonumber & = \rp{G S_\textrm{X}}{c^2 r_\textrm{X}^5} \qua{\ton{-\kx r_\textrm{X}^2 + 3 \kx {x^2_\textrm{X}} + 3 \ky{x_\textrm{X}}{y_\textrm{X}} + 3 \kz{x_\textrm{X}}{z_\textrm{X}}} \cos\Omega + \right.\\ \nonumber \\
&+ \left.\ton{-\ky r_\textrm{X}^2 + 3 \kx{x_\textrm{X}}{y_\textrm{X}} + 3 \ky {y_\textrm{X}}^2 + 3 \kz {y_\textrm{X}} {z_\textrm{X}}} \sin\Omega}, \\ \nonumber \\
A_T^\textrm{GM} \nonumber & = -\rp{G S_\textrm{X} \csc I}{c^2 r_\textrm{X}^5} \grf{\sin I\qua{-3 \ton{\kx{x_\textrm{X}}+ \ky {y_\textrm{X}}} {z_\textrm{X}} + \kz \ton{r_\textrm{X}^2 - 3 {z_\textrm{X}}^2}}  + \right.\\ \nonumber \\
\nonumber &+\left.  \cos I\qua{\ky \ton{r_\textrm{X}^2 - 3 {y_\textrm{X}}^2} - 3  \ton{\kx{x_\textrm{X}}+ \kz {z_\textrm{X}}}{y_\textrm{X}}} \cos\Omega + \right.\\ \nonumber \\
&+\left. \cos I\qua{\kx \ton{-r_\textrm{X}^2 + 3 {x^2_\textrm{X}}} + 3 \ky{x_\textrm{X}}{y_\textrm{X}} + 3 \kz{x_\textrm{X}}{z_\textrm{X}}} \sin\Omega}, \\ \nonumber \\
A_N^\textrm{GM} \nonumber & = \rp{G S_\textrm{X} \csc I}{c^2 r_\textrm{X}^5} \grf{\qua{\ky \ton{r_\textrm{X}^2 - 3 {y_\textrm{X}}^2} - 3 {y_\textrm{X}} \ton{\kx{x_\textrm{X}}+ \kz {z_\textrm{X}}}} \cos\Omega + \right.\\ \nonumber \\
&+\left. \qua{\kx \ton{-r_\textrm{X}^2 + 3 {x^2_\textrm{X}}} + 3 \ky{x_\textrm{X}}{y_\textrm{X}} + 3 \kz{x_\textrm{X}}{z_\textrm{X}}} \sin\Omega}.
\end{align}
They have to be inserted into the right-hand-sides of the standard Gauss equations for the variation of the orbital elements \citep{Bertotti03,2011rcms.book.....K,2014grav.book.....P} which, finally, are doubly averaged with respect to $P_\textrm{b},~P_\textrm{b}^\textrm{X}$ in the same way as previously described for the disturbing potential of \rfr{USS}.

\begin{figure*}
\begin{center}
\centerline{
\vbox{
\begin{tabular}{cc}
\epsfxsize= 8.0 cm\epsfbox{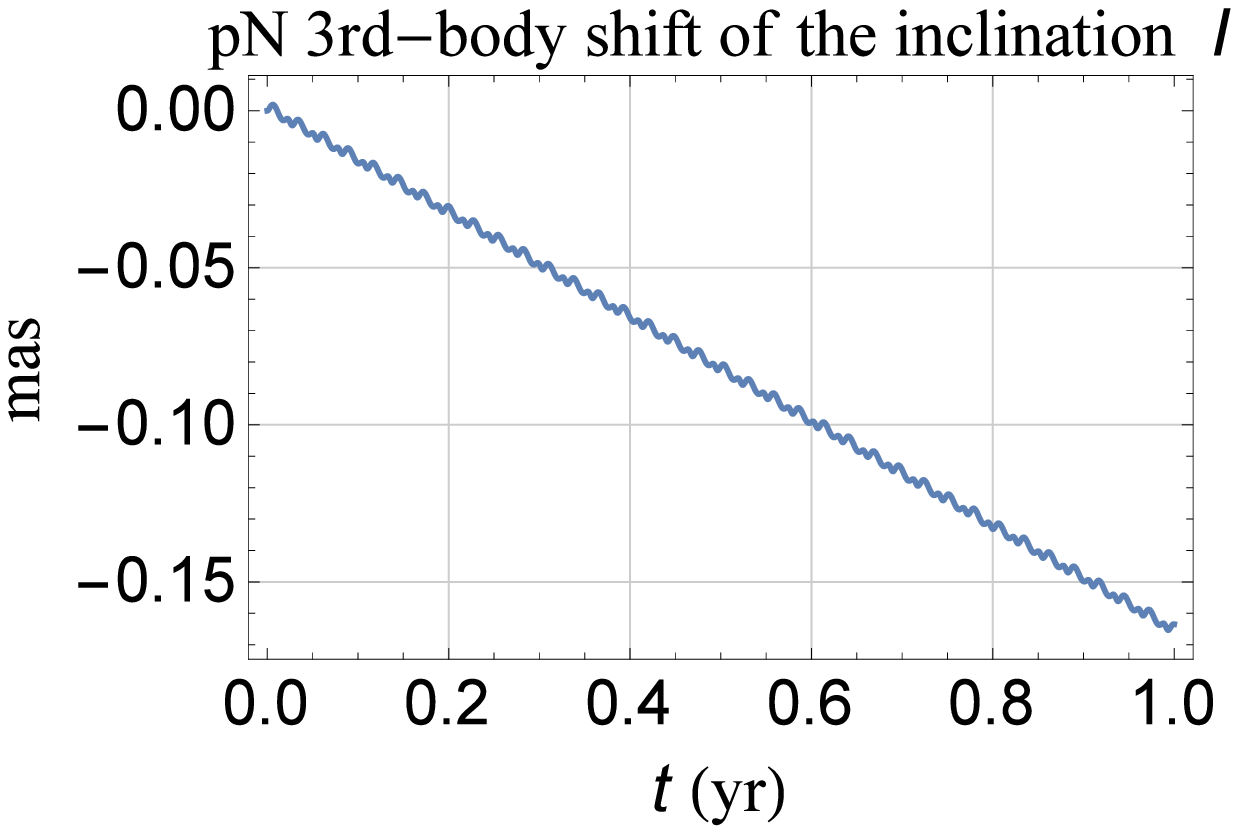}\\
\epsfxsize= 8.0 cm\epsfbox{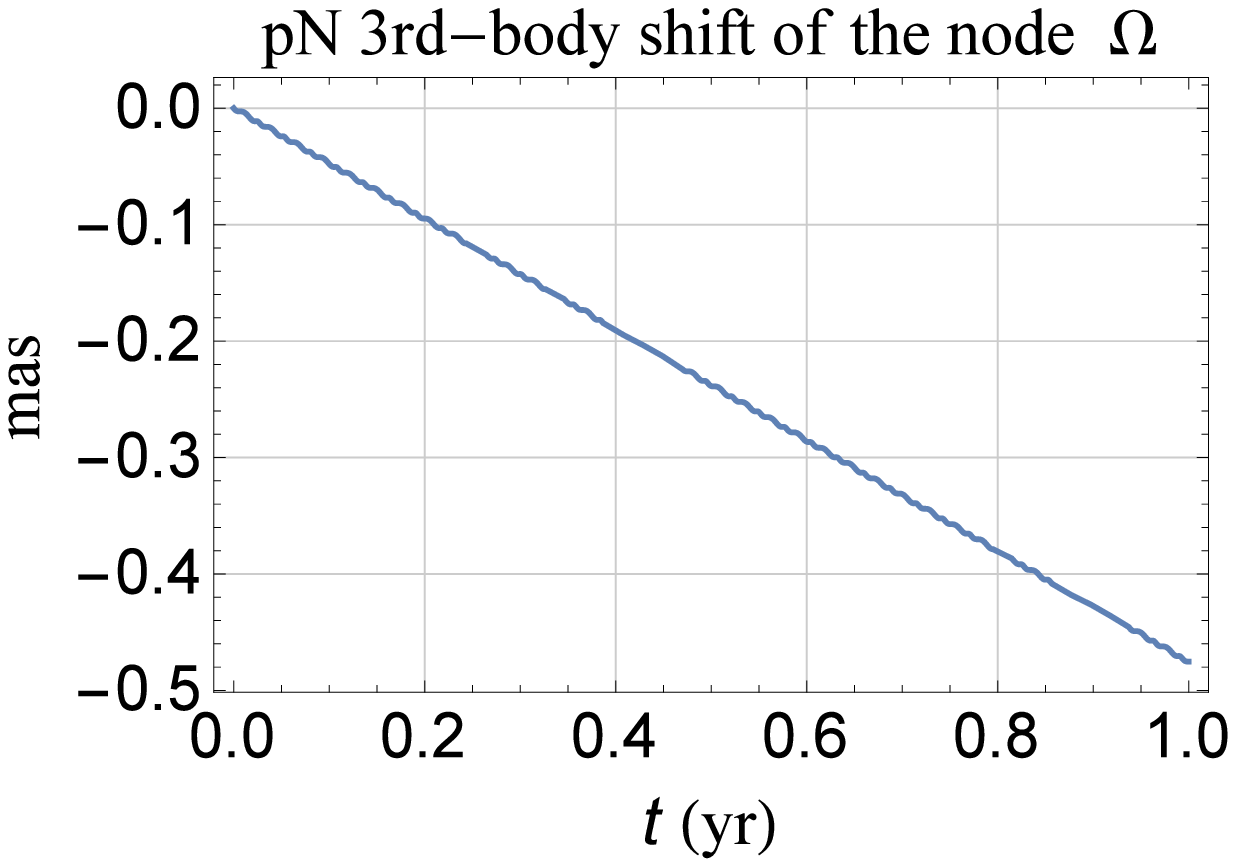}\\
\epsfxsize= 8.0 cm\epsfbox{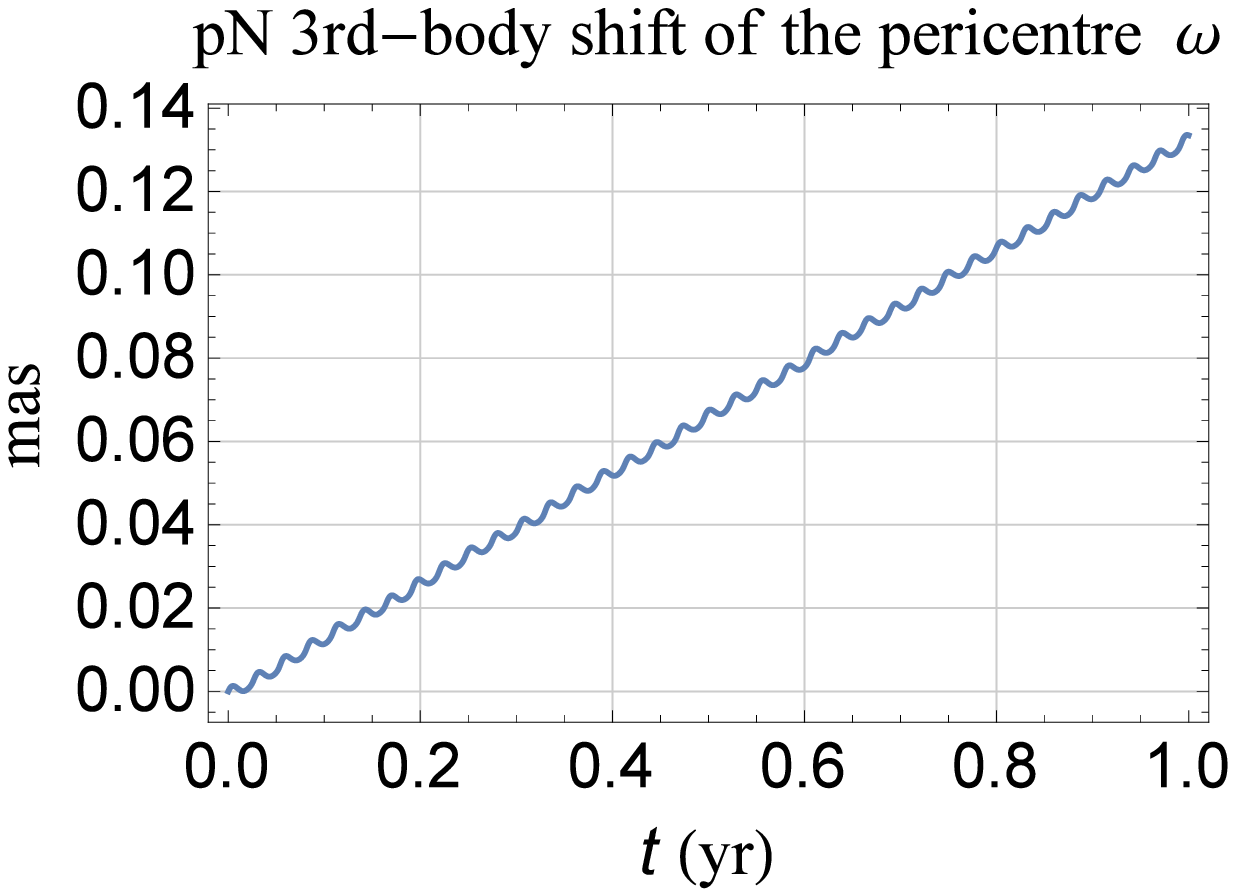}\\
\end{tabular}
}
}
\caption{Numerically computed time series of the post-Newtonian gravitomagnetic 3rd-body shifts experienced by the  inclination $I$, node $\Omega$ and pericentre $\omega$ of a fictitious test particle moving around a Callisto-like primary $M$ which, in turn, orbits a Jupiter-type 3rd-body X. They were obtained by numerically integrating the equations of motion of the orbiter in Cartesian rectangular coordinates referred to the Earth's mean equator at the epoch J2000.0 with and without \rfr{ASS}. Both runs shared the same set of arbitrary initial conditions for the probe $\Pb = 10.07~\textrm{d},~e_0 = 0.3,~I_0 = 80\deg,~\Omega_0 = 230\deg,~\omega_0 = 40\deg,~f_0 = 50\deg$ and the primary; as far as the motion of $M$ with respect to X is concerned, the initial state vector of the Callisto-Jupiter relative motion was adopted from the database JPL HORIZONS
(https://ssd.jpl.nasa.gov/?horizons). For each Keplerian orbital element, its time series calculated  from the purely Newtonian run was subtracted from that obtained from the post-Newtonian integration in order to obtain the signatures displayed here.
The resulting shifts, in  $\masy$, agree with the analytically computed ones in Equations~(\ref{Idot})~to~(\ref{odot}).
}\label{fig0}
\end{center}
\end{figure*}

We successfully checked our analytical results of Equations~(\ref{adot})~to~(\ref{ocoef}) as follows. We considered a fictitious system $\mathcal{S}$ orbiting a Jupiter-like body X along the same orbit of Callisto, whose mass was assumed for the particle's primary $M$, and numerically integrated its equations of motion  over a time span much longer than $\Pb,~\Pb^\textrm{X}$ with and without the post-Newtonian gravitomagnetic acceleration of \rfr{ASS}; both the integrations, which assumed a purely Keplerian motion of $\mathcal{S}$ about the fictitious body X, shared the same initial conditions for the test particle and its primary. For X, the same physical properties of Jupiter were assumed, including the size and the orientation of its angular momentum $\bds S$. As a result, numerically produced times series of the orbital elements of the imaginary probe were produced by subtracting the purely Newtonian ones from those obtained by including also \rfr{ASS} in the equations of motion; they are displayed in Figure~\ref{fig0}. It turned out that the resulting numerically calculated post-Newtonian gravitomagnetic 3-body orbital shifts agree with those computed by means of the analytical formulas of Equations~(\ref{adot})~to~(\ref{ocoef}).
%
\section{Some potentially interesting astronomical scenarios}\lb{encel}
For the sake of simplicity, we will consider a circular orbit ($e=0$) for the test particle motion around $M$.

In the case of a hypothetical orbiter of the Kronian natural satellite Enceladus in the external field of Saturn, \rfrs{Idot}{Odot} and \rfrs{Icoef}{Ocoef}, referred to the mean Earth's equator at the reference epoch J2000.0 as reference $\grf{x,~y}$ plane, yield
\begin{align}
\dot I & = A_\textrm{eq}\sin\ton{\Omega + \varphi_\textrm{eq}},\\ \nonumber \\
\dot\Omega &= -49.9\masy + \cot I~A_\textrm{eq}\cos\ton{\Omega + \varphi_\textrm{eq}},
\end{align}
with
\begin{align}
A_\textrm{eq} & = -5.7\masy, \\ \nonumber \\
\varphi_\textrm{eq} &= 49.4\deg.
\end{align}
Instead, if the mean ecliptic at the reference epoch J2000.0 is adopted as reference $\grf{x,~y}$ plane, we have
 \begin{align}
\dot I & = A_\textrm{ecl}\sin\ton{\Omega + \varphi_\textrm{ecl}},\\ \nonumber \\
\dot\Omega &= -34.0\masy + \cot I~A_\textrm{ecl}\cos\ton{\Omega + \varphi_\textrm{ecl}},
\end{align}
with
\begin{align}
A_\textrm{ecl} & = 23.9\masy, \\ \nonumber \\
\varphi_\textrm{ecl} &= 10.4\deg.
\end{align}

\begin{figure*}
\begin{center}
\centerline{
\vbox{
\begin{tabular}{c}
\epsfbox{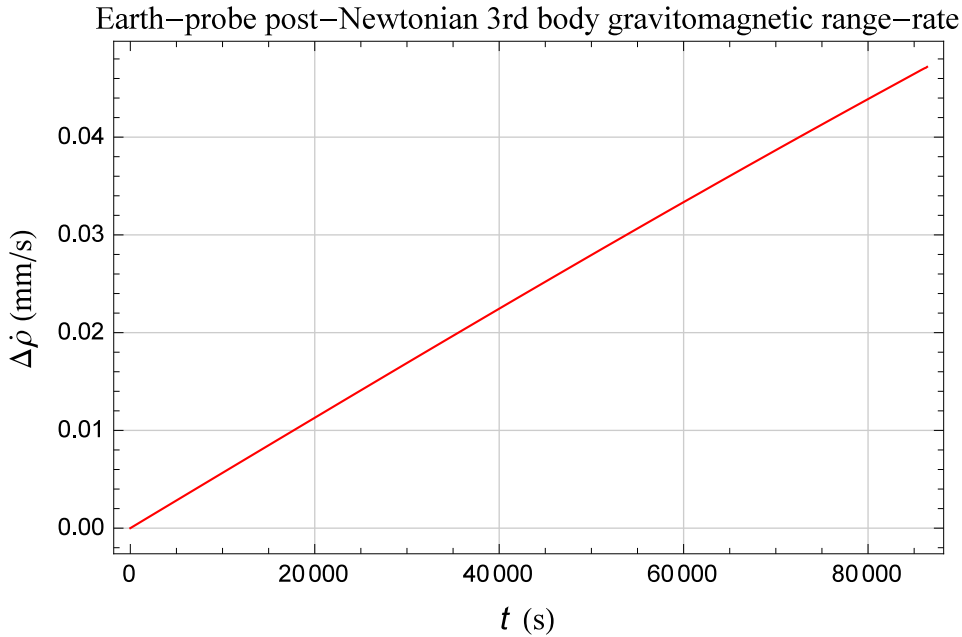}\\
\end{tabular}
}
}
\caption{Numerically produced Earth-probe range-rate shift $\Delta\dot\rho\ton{t}$ due to the post-Newtonian gravitomagnetic 3rd-body acceleration of \rfr{ASS}. We numerically integrated the solar system barycentric equations of motion in Cartesian rectangular coordinates of the Earth, Jupiter, its Galilean moons and a fictitious test particle orbiting Europa over 1 d. In both runs, sharing the same initial conditions for all the existing natural bodies retrieved from the database JPL HORIZONS (https://ssd.jpl.nasa.gov/?horizons) at the arbitrary epoch of midnight of 1st January 2030, we modeled the mutual attractions among all the planets and the satellites involved to the Newtonian level, with the exception of \rfr{ASS} which was added to the other classical gravitational pulls felt by the orbiter in one of the runs. Then, we numerically calculated two range-rate time series, and subtracted the purely Newtonian one from that including also \rfr{ASS}. The orbital configuration adopted for the spacecraft, referred to Europa,  was $a_0 = 3.55~R,~e_0 = 0.69,~I_0 = 100\deg,~\Omega_0 = 90\deg,~\omega_0 = 40\deg,~f_0 = 50\deg$, where $R$ is the radius of Jovian moon.
}\label{fig1}
\end{center}
\end{figure*}

By looking at a putative orbiter of the Jovian natural satellite Europa in the external field of Jupiter, we have
\begin{align}
\dot I & = A_\textrm{eq}\sin\ton{\Omega + \varphi_\textrm{eq}},\\ \nonumber \\
\dot\Omega &= -9.9\masy + \cot I~A_\textrm{eq}\cos\ton{\Omega + \varphi_\textrm{eq}},
\end{align}
with
\begin{align}
A_\textrm{eq} & = 4.8\masy, \\ \nonumber \\
\varphi_\textrm{eq} &= 2.9\deg, \\ \nonumber \\
\end{align}
and
\begin{align}
\dot I & = A_\textrm{ecl}\sin\ton{\Omega + \varphi_\textrm{ecl}},\\ \nonumber \\
\dot\Omega &= -11.0\masy + \cot I~A_\textrm{ecl}\cos\ton{\Omega + \varphi_\textrm{ecl}},
\end{align}
with
\begin{align}
A_\textrm{ecl} & = 0.3\masy, \\ \nonumber \\
\varphi_\textrm{ecl} &= 31.0\deg.
\end{align}
We considered just Enceladus and Europa because they are of great planetological interest in view of the possible habitability of their oceans beneath their
icy crusts \citep{2017AcAau.131..123L}. As a consequence, they are the natural targets of several concept studies and proposals for dedicated missions to them, including also orbiters \citep{2008AIPC..969..388R,EO2010, 2016AdSpR..58.1117M,2018Icar..314...35V,2018AcAau.143..285S}.
Since, at present, sending a spacecraft to Europa seems more likely than to Enceladus, as it can be learnt at https://europa.nasa.gov/about-clipper/overview/ and http://sci.esa.int/juice/ on the Internet, we investigated in a little more detail this potentially appealing Jovian scenario, even if it is not said that the actually approved missions will finally involve the use of an orbiter. In such kind of endeavours, the observable quantity is typically the Earth-probe range-rate $\dot\rho$, whose accuracy for, e.g., the ongoing mission Juno \citep{2017SSRv..213....5B} around Jupiter is $\upsigma_{\dot\rho}\simeq 0.015~\textrm{mm~s}^{-1}$ \citep{2018Natur.555..220I}. Figure~\ref{fig1} shows the numerically simulated Earth-spacecraft range-rate signature due to the post-Newtonian gravitomagnetic 3-body acceleration of \rfr{ASS} for a generic orbital configuration of the hypothesized orbiter. To produce it, we numerically integrated the equations of motion in Cartesian rectangular coordinates of the Earth, Jupiter, its Galilean moons and a fictitious test particle orbiting Europa over 1 d. In both runs, sharing the same initial conditions retrieved from the database JPL HORIZONS (https://ssd.jpl.nasa.gov/?horizons) at the arbitrary epoch of midnight of 1st January 2030, we modeled the mutual attractions among all the bodies involved to the Newtonian level, with the exception of \rfr{ASS} which was added to the other classical gravitational pulls felt by the probe in one of the runs. Then, we numerically calculated two range-rate time series, and subtracted the purely Newtonian one from that including also the post-Newtonian gravitomagnetic acceleration.
It can be noted that, for the orbital configuration chosen, the range-rate relativistic signature $\Delta\dot\rho$ reaches the $0.05~\textrm{mm~s}^{-1}$ level after just 1 d. Thus, the scenario considered seems worth of further, dedicated analyses investigating the actual measurability of \rfr{ASS} in a realistic error budget analysis and mission proposal. It should take into account several concurring perturbations of gravitational and non-gravitational nature, and also several technological and engineering issues.

In the case of an artificial satellite orbiting a planet in the field of the Sun, the effects are much smaller.
For an Earth's spacecraft, we have
\begin{align}
\dot I & = A_\textrm{eq}\sin\ton{\Omega + \varphi_\textrm{eq}},\\ \nonumber \\
\dot\Omega &= -0.2\muasy + \cot I~A_\textrm{eq}\cos\ton{\Omega + \varphi_\textrm{eq}},
\end{align}
with
\begin{align}
A_\textrm{eq} & = 0.1\muasy, \\ \nonumber \\
\varphi_\textrm{eq} & = 9.13\deg,
\end{align}
and
\begin{align}
\dot I & = A_\textrm{ecl}\sin\ton{\Omega + \varphi_\textrm{ecl}},\\ \nonumber \\
\dot\Omega &= -0.3\muasy + \cot I~A_\textrm{ecl}\cos\ton{\Omega + \varphi_\textrm{ecl}},
\end{align}
with
\begin{align}
A_\textrm{ecl} & = 0.02\muasy, \\ \nonumber \\
\varphi_\textrm{ecl} & = 104.2\deg.
\end{align}
For a probe orbiting Mercury one gets
\begin{align}
\dot I & = A_\textrm{eq}\sin\ton{\Omega + \varphi_\textrm{eq}},\\ \nonumber \\
\dot\Omega &= -4.3\muasy + \cot I~A_\textrm{eq}\cos\ton{\Omega + \varphi_\textrm{eq}},
\end{align}
with
\begin{align}
A_\textrm{eq} & = -2.5\muasy, \\ \nonumber \\
\varphi_\textrm{eq} & = 171.3\deg,
\end{align}
and
\begin{align}
\dot I & = A_\textrm{ecl}\sin\ton{\Omega + \varphi_\textrm{ecl}},\\ \nonumber \\
\dot\Omega &= -5\muasy + \cot I~A_\textrm{ecl}\cos\ton{\Omega + \varphi_\textrm{ecl}},
\end{align}
with
\begin{align}
A_\textrm{ecl} & = -0.6\muasy, \\ \nonumber \\
\varphi_\textrm{ecl} & = 144.6\deg.
\end{align}
\section{Summary and overview}\lb{conclu}
In the weak-field and slow-motion approximation of general relativity, we analytically worked out the post-Newtonian gravitomagnetic long-term rates of change of the relevant Keplerian orbital elements of a test particle orbiting a primary $M$ at distance $r$ from it which, in turn, moves in the external spacetime deformed by the mass-energy currents of the spin angular momentum ${\bds S}_\textrm{X}$ of a distant ($r_\textrm{X}\gg r$) 3rd body X with mass $M_\textrm{X}\gg M$. We did not assume any preferred orientation for the spin axis ${\bds{\hat{S}}}_\textrm{X}$ of the external body; moreover, we did not make simplifying assumptions pertaining the orbital configurations of both the $M$'s satellite and of $M$ itself in its motion around $M_\textrm{X}$. Thus, our calculation have a general validity, being applicable to arbitrary astronomical systems of potential interest.
It turns out that, by doubly averaging the perturbing potential employed in the calculation with respect to the orbital periods $\Pb,~\Pb^\textrm{X}$ of both $M$ and $M_\textrm{X}$, the semimajor axis $a$ and the eccentricity $e$ do not experience long-term variations, contrary to the inclination $I$ of the orbital plane, the longitude of the ascending node $\Omega$ and the argument of pericenter $\omega$. While the gravitomagnetic rates $\dot I$ and $\dot \omega$ are harmonic signatures characterized by the frequency of the possible variation of the node $\Omega$, induced by other dominant perturbations like, e.g., the Newtonian quadrupole mass moment of the satellite's primary $M$, the gravitomagnetic node rate $\dot\Omega$ exhibits also a secular trend in addition to a harmonic component with the frequency of the node itself. A numerical integration of the equations of motion of a fictitious 3-body system made of a distant spinning body with the same physical properties of Jupiter, a primary with the same orbital and physical characteristics of Callisto and a test particle orbiting it confirms our analytical results.

The Sun's angular momentum  exerts very small effects on spacecraft orbiting Mercury ($\simeq 1\muasy$) and the Earth ($\sim 0.1\muasy$). Instead, the angular momenta of the gaseous giant planets like Jupiter and Saturn may induce much larger perturbations of the orbital motions of hypothetical anthropogenic orbiters of some of their major natural moons like, e.g., Europa ($\lesssim 10\masy$) and Enceladus ($\lesssim 50\masy$). Such natural satellites have preeminent interest in planetology making them ideal targets for future, dedicated spacecraft-based missions which may be opportunistically exploited to attempt to measure such relativistic effects as well. In the case of Europa, for whose exploration there are  already approved missions by NASA and ESA, a preliminary numerical simulation of the signature induced by the post-Newtonian gravitomagnetic 3-body effect of interest on the range-rate of a putative orbiter shows that, for certain orbital configurations, its magnitude can become larger than the present-day accuracy $\upsigma_{\dot \rho}=0.015~\textrm{mm~s}^{-1}$ of the current Juno mission around Jupiter after 1 d.
\bibliography{Gclockbib,semimabib,PXbib}{}

\begin{appendices}
\section{Notations and definitions}\lb{appena}
Here, some basic notations and definitions pertaining the restricted two-body system $\mathcal{S}$ moving in the external field of the distant 3rd-body X considered in the text are presented. For the numerical values of some of them, see Tables~\ref{tavola0}~and~\ref{tavola1}.
\begin{description}
\item[] $G:$ Newtonian constant of gravitation
\item[] $c:$ speed of light in vacuum
\item[] $\epsilon:$ mean obliquity
\item[] $M_\textrm{X}:$ mass of the distant 3rd-body X (a star like the Sun or a planet like, e.g., Jupiter or Saturn)
\item[] $\mu_\textrm{X}\doteq GM_\textrm{X}:$ gravitational parameter of the 3rd-body X
\item[] $S_\textrm{X}:$ magnitude of the angular momentum of the 3rd-body X
\item[] ${\bds{\hat{S}}}_\textrm{X} = \grf{\kx,~\ky,~\kz}:$ spin axis of the 3rd-body X in some coordinate system
\item[] $\alpha_\textrm{X}:$ right ascension (RA) of the 3rd-body's spin axis
\item[] $\delta_\textrm{X}:$ declination (DEC) of the 3rd-body's spin axis
\item[] $\kx^\textrm{eq}=\cos\delta_\textrm{X}\cos\alpha_\textrm{X}:$ component of the 3rd-body's spin axis w.r.t. the reference $x$ axis of an equatorial coordinate system
\item[] $\ky^\textrm{eq}=\cos\delta_\textrm{X}\sin\alpha_\textrm{X}:$ component of the 3rd-body's spin axis w.r.t. the reference $y$ axis of an equatorial coordinate system
\item[] $\kz^\textrm{eq}=\sin\delta_\textrm{X}:$ component of the 3rd-body's spin axis w.r.t. the reference $z$ axis of an equatorial coordinate system
\item[] ${\bds r}_\textrm{X}:$ position vector towards the 3rd-body X
\item[] $r_\textrm{X}:$ distance of $\mathcal{S}$ to the 3rd-body X
\item[] ${\bds{\hat{r}}}_\textrm{X}\doteq {\bds r}_\textrm{X}/r_\textrm{X}:$ versor of the position vector towards the 3rd-body X
\item[] $a_\textrm{X}:$  semimajor axis of the orbit about the 3rd-body X
\item[] $\nk^\textrm{X}\doteq \sqrt{\mu_\textrm{X}/\aX^3}:$ mean motion of the orbit about the 3rd-body X
\item[] $P_\textrm{b}^\textrm{X}\doteq 2\uppi/\nk^\textrm{X}:$ orbital period of the orbit about the 3rd-body X
\item[] $e_\textrm{X}:$  eccentricity of the orbit about the 3rd-body X
\item[] $I_\textrm{X}:$  inclination of the orbital plane of orbit about the 3rd-body X to the reference $\grf{x,~y}$ plane of some coordinate system
\item[] $\Omega_\textrm{X}:$  longitude of the ascending node  of the orbit about the 3rd-body X referred to the reference $\grf{x,~y}$ plane of some coordinate system
\item[] $M:$ mass of the primary (planet or planetary natural satellite) orbited by the test particle and moving in the external field of the 3rd-body X
\item[] $\mu\doteq GM:$ gravitational parameter of the primary orbited by the test particle and moving in the external field of the 3rd-body X
\item[] $R:$ radius of the primary (planet or planetary natural satellite) orbited by the test particle and moving in the external field of the 3rd-body X
\item[] $\bds S:$ angular momentum of the primary
\item[] $J_\ell,~\ell=2,3,4~:$ zonal multipole moments of the classical gravitational potential of the primary
\item[] $\bds r:$ position vector of the test particle with respect to its primary
\item[] $r:$ magnitude of the position vector of the test particle
\item[] $\bds v:$ velocity vector of the test particle
\item[] $\bds L\doteq \bds r\bds\times\bds v:$ orbital angular momentum per unit mass of the test particle
\item[] $a:$  semimajor axis of the test particle's orbit
\item[] $\nk \doteq \sqrt{\mu/a^3}:$  Keplerian mean motion of the test particle's orbit
\item[] $\Pb\doteq 2\uppi/\nk:$ orbital period of the test particle's orbit
\item[] $e:$  eccentricity of the test particle's orbit
\item[] $f:$ true anomaly of the test particle's orbit
\item[] $I:$  inclination of the orbital plane of the test particle's orbit to the reference $\grf{x,~y}$ plane of some coordinate system
\item[] $\Omega:$  longitude of the ascending node  of the test particle's orbit referred to the reference $\grf{x,~y}$ plane of some coordinate system
\end{description}
\section{Tables}\lb{appenb}
\begin{table*}
\caption{Relevant physical and orbital parameters for Saturn, Jupiter, Enceladus and Europa. Most of the reported values come from \citet{2003AJ....126.2687S,2007CeMDA..98..155S,2010ITN....36....1P} and references therein. The source for the orbital elements referred to either the mean ecliptic (ecl) at the reference epoch J2000.0 or the mean Earth's equator (eq) at the same epoch  is the freely consultable database JPL HORIZONS on the Internet at https://ssd.jpl.nasa.gov/?horizons from which they were retrieved by choosing the time of writing this paper as input epoch.
}\lb{tavola0}
\begin{center}
\begin{tabular}{|l|l|l|}
  \hline
Parameter  & Units & Numerical value \\
\hline
$G$ & $\textrm{kg~m}^3~\textrm{s}^{-2}$ & $6.67259\times 10^{-11} $\\
$c$ & $\textrm{m~s}^{-1}$ & $2.99792458\times 10^8$\\
\hline
$S_{\saturn}$ & $\textrm{kg~m}^2$~$\textrm{s}^{-1}$ & $1.4\times 10^{38}$\\
$\alpha_{\saturn}$ & $\textrm{deg}$ & $40.59$\\
$\delta_{\saturn}$ & $\textrm{deg}$ & $83.54$\\
$a_\textrm{Enc}$ & $\textrm{km}$ & $237,948$\\
$e_\textrm{Enc}$ & $-$ & $0.0047$\\
$I_\textrm{Enc}^\textrm{eq}$ & $\textrm{deg}$ & $6.475336858877378$ \\
$I_\textrm{Enc}^\textrm{ecl}$ & $\textrm{deg}$ & $28.06170970578348$ \\
$\Omega_\textrm{Enc}^\textrm{eq}$ & $\textrm{deg}$ & $130.5900992493321$ \\
$\Omega_\textrm{Enc}^\textrm{ecl}$ & $\textrm{deg}$ & $169.5108697290241$ \\
$\Pb^\textrm{Enc}$ & d & $1.370218$\\
\hline
$S_{\jupiter}$ & $\textrm{kg~m}^2$~$\textrm{s}^{-1}$ & $6.9\times 10^{38}$\\
$\alpha_{\jupiter}$ & $\textrm{deg}$ & $268.05$\\
$\delta_{\jupiter}$ & $\textrm{deg}$ & $64.49$\\
$a_\textrm{Eur}$ & $\textrm{km}$ & $671,034 $\\
$e_\textrm{Eur}$ & $-$ & $0.0094$\\
$I_\textrm{Eur}^\textrm{eq}$ & $\textrm{deg}$ & $25.88280598312641$ \\
$I_\textrm{Eur}^\textrm{ecl}$ & $\textrm{deg}$ & $1.790876103183550$ \\
$\Omega_\textrm{Eur}^\textrm{eq}$ & $\textrm{deg}$ & $357.4169659423443$ \\
$\Omega_\textrm{Eur}^\textrm{ecl}$ & $\textrm{deg}$ & $332.6268549691798$ \\
$\Pb^\textrm{Eur}$ & d & $3.551810$\\
\hline
\end{tabular}
\end{center}
\end{table*}
\begin{table*}
\caption{Relevant physical and orbital parameters used in the text for the Sun, Mercury and the Earth. Most of the reported values come from \citet{2003AJ....126.2687S,2007CeMDA..98..155S,2010ITN....36....1P} and references therein. The source for the orbital elements referred to either the mean ecliptic (ecl) at the reference epoch J2000.0 or the mean Earth's equator (eq) at the same epoch  is the freely consultable database JPL HORIZONS on the Internet at https://ssd.jpl.nasa.gov/?horizons from which they were retrieved by choosing the time of writing this paper as input epoch.
}\lb{tavola1}
\begin{center}
\begin{tabular}{|l|l|l|}
  \hline
Parameter  & Units & Numerical value \\
\hline
$G$ & $\textrm{kg}^{-1}~\textrm{m}^3~\textrm{s}^{-2}$ & $6.67259\times 10^{-11} $\\
$c$ & $\textrm{m~s}^{-1}$ & $2.99792458\times 10^8$\\
\hline
$S_{\odot}$ & $\textrm{kg~m}^2$~$\textrm{s}^{-1}$ & $1.90\times 10^{41}$\\
$\alpha_{\odot}$ & $\textrm{deg}$ & $286.13$\\
$\delta_{\odot}$ & $\textrm{deg}$ & $63.87$\\
\hline
$a_{\mercury}$ & $\textrm{au}$ & $0.3870982252717257$\\
$e_{\mercury}$ & $-$ & $0.2056302512089075$\\
$I_{\mercury}^\textrm{eq}$ & $\textrm{deg}$ & $28.55225598038233$ \\
$I_{\mercury}^\textrm{ecl}$ & $\textrm{deg}$ & $7.005014199657344$ \\
$\Omega_{\mercury}^\textrm{eq}$ & $\textrm{deg}$ & $10.98794759075666$ \\
$\Omega_{\mercury}^\textrm{ecl}$ & $\textrm{deg}$ & $48.33053756455964$ \\
$\Pb^{\mercury}$ & yr & $0.2408467$\\
\hline
$a_\oplus$ & $\textrm{au}$ & $0.9992521882390240$\\
$e_\oplus$ & $-$ & $0.01731885059206812$\\
$I_\oplus^\textrm{eq}$ & $\textrm{deg}$ & $23.43903457134406$ \\
$I_\oplus^\textrm{ecl}$ & $\textrm{deg}$ & $2.669113820737183\times 10^{-4}$ \\
$\Omega_\oplus^\textrm{eq}$ & $\textrm{deg}$ & $1.852352676284691\times 10^{-4}$ \\
$\Omega_\oplus^\textrm{ecl}$ & $\textrm{deg}$ & $163.9752443600624$ \\
$\Pb^\oplus$ & yr & $1.0000174$\\
\hline
\end{tabular}
\end{center}
\end{table*}
\clearpage
\end{appendices}
\end{document}